\documentclass[aps,prl,twocolumn,superscriptaddress]{revtex4-1}
\usepackage{amssymb, graphicx, hyperref, subcaption, color, hyperref}
\usepackage[intlimits]{amsmath}
\usepackage[english]{babel}
\begin{document}

\title{Majorana flipping of quarkonium spin states in transient magnetic field }

\author{Nirupam Dutta}
\email[E-mail:]{nirupamdu@gmail.com}
\affiliation{Theoretical Physics Division, Variable Energy Cyclotron Centre,
1/AF, Bidhan Nagar, 
Kolkata 700064, India}

\author{Surasree Mazumder}
\email[E-mail:]{surasree.mazumder@gmail.com}
\affiliation{Saha Institute of Nuclear Physics, 1/AF, Bidhan Nagar, Kolkata 700064, India}

\date{\today}

\begin{abstract}

We demonstrate that spin flipping transitions occur between various quarkonium spin states due to 
transient magnetic field produced in non central heavy ion collisions (HICs). The 
inhomogeneous nature of the magnetic field results in \textit{non adiabatic evolution} of  
(spin)states of quarkonia moving inside the transient magnetic environment. Our calculations 
explicitly show that the consideration of azimuthal inhomogeneity gives rise to dynamical mixing 
between different spin states owing to  Majorana spin flipping. Notably, this effect of  non-adiabaticity
is novel and distinct from previously predicted mixing of the singlet and one of the triplet states 
of quarkonia in the presence of a static and homogeneous magnetic field.

\end{abstract}

\maketitle

The study of magnetic field in heavy ion collisions(HICs) has become an 
exciting trend in recent years.
The existence of a strong time varying magnetic field is theoretically 
conjectured \cite{skokov2009,kharzeev2007,rafelski1975,voronyuk2011} and is  awaiting experimental verification. 
Physicists are investing a lot of effort and interest to observe chiral magnetic effect \cite{kharzeev2007,kharzeev2006,bloczynski2012}. 
Not only that, many other observables, too, will be affected should the magnetic field
be produced. Therefore, it is quite natural to see how this magnetic field influences 
heavy quarkonia which, by themselves, constitute one of the major probes for the medium formed in 
high energy nucleus-nucleus collisions. It is obvious that the external magnetic field 
 affects both the spatial and spin degrees of freedom of quarkonia. It is shown in some literature 
\cite{marasinghe2011,bonati2015,guo2015,Tuchin2011} how the presence of such a field impacts on
the formation and dissociation of heavy quark-anti quark bound states. As far as the interaction 
between spin and magnetic field is concerned, the analogy with positronium atom \cite{karshenboim2003,filip2013,michael2013} is adopted in 
studying the Zeeman splitting and spin mixing of quarkonium spin states. It shows that, for quarkonia
also, there is spin mixing between ortho and para states, such as between 
$1S$ triplet ($J/\Psi$ and $\eta_c$) and singlet ($\Upsilon$ and $\eta_b$) states of
charmonium and bottomonium respectively. This realisation is entirely based on the consideration of a 
homogeneous and constant magnetic field although a time dependent field has sometimes been considered
\cite{yang2011}. The nature of spin mixing can radically change if we introduce an inhomogeneous
magnetic field instead. Besides, the knowledge of the exact nature of this magnetic field 
is highly speculative till date. There are a few studies indicating that the field
only lasts for a narrow span of time \cite{tuchin2013,skokov2009}, whereas, other investigations suggest that it can 
linger on for a longer time \cite{ajit2017} due to the presence of the quark gluon plasma(QGP) medium formed
after collision. Ergo, we decide to tackle the problem of quarkonium spin mixing from a general standpoint 
where the magnetic field varies both spatially and temporally. Notably, the centre of mass of quarkonia
is not static, rather, it moves in the inhomogeneous magnetic environment. As the 
quarkonium  travels from one space point to another, it  experiences red a varying interaction
potential owing to the spatially changing magnetic field. Hence, the Hamiltonian of the system is 
changing with time. This time dependence becomes explicit when observed from the co-moving frame
of quarkonia. Such a time dependent Hamiltonian invites us
to check whether the nature of evolution of quarkonium spin states is adiabatic or non-adiabatic. 
Non-adiabaticity can cause spin flipping transitions (Majorana flipping) between 
different spin states. This sort of effect of non adiabaticity 
 has, thus far, not been considered to the best of our knowledge. The time rate of change of the magnetic field,
when observed from the rest frame of quarkonia, depends  on the speed of quarkonia
inside the inhomogeneous magnetic field. If the time scale of this varying field is 
much much smaller than that of the evolution of quarkonium spin states, the system 
will evolve adiabatically. On the other hand,  if the magnetic field produced in HICs is rapidly decaying, 
it will guarantee the existence of a weak field regime where the Larmor frequency,
being proportional to the field strength, is also negligibly small. As the time scale of evolution of 
quarkonium spin sates is inversely proportional to Larmor frequency, $\omega$, the weak
field or small $\omega$ leads to a non-adiabatic evolution. In that case,
one cannot naively hold fast to adiabaticity, rather, intricacies of non-adiabaticity
may come into play.

In this article, the necessary and sufficient conditions for the occurrence of adiabatic 
evolution are discussed. To this end, the spin flipping transition probabilities 
for different spin states have been estimated by solving the 
Schr\"odinger equation with time dependent Hamiltonian. 
It has already been mentioned that previous works with a homogeneous
field observed mixing between one of the triplet states and the singlet state of 
quarkonia \cite{michael2013}. \textit{Contrarily, considering an azimuthal inhomogeneity of magnetic field, we have explicitly 
witnessed the possibility of dynamical spin mixing, not only between two states,
but amongst all possible states}.

\section*{Adiabatic and Non-adiabatic spin mixing:}

The Hamiltonian of the system can be written as a summation of unperturbed part ($H_{0}$) and 
interaction part ($H_I$) due to the magnetic field. $H_{0}$ only contains the spatial degrees of freedom
because spin spin interaction has not been considered in the present work. 
Our point of interest is  the spin-field interaction in $H_I$ and any other
term that might appear in $H$ \cite{michael2013,bonati2015} has not been taken into account here. So,
\begin{align}
H = H_{0}-\vec{\mu} \cdot \vec{B}~. \label{eq:interaction}
\end{align}
The quantity $\vec{\mu} $ is the magnetic moment of the bound $q\bar{q}$ pair.
Singling out the interaction part, one can write
\begin{align}
H_I=& -(\vec{\mu}_Q+\vec{\mu}_{\bar{Q}})\cdot \vec{B}\nonumber\\
=& -(g_Q\mu_Q\vec{S}_Q+g_{\bar{Q}}\mu_{\bar{Q}}\vec{S}_{\bar{Q}})\cdot \vec{B}\nonumber\\
=& -g\mu_Q(\vec{S}_Q-\vec{S}_{\bar{Q}})\cdot \vec{B} \label{moment}
\end{align}
Here, $g$ is the gyromagnetic ratio where, $g=g_Q=-g_{\bar{Q}}$, $\vec{S}$ is the spin 
and $\mu_Q$ is the quark magneton  given
by $Q/2m_Q$, $m_Q$ being the mass of quark/antiquark.
To start with, let us take the applied magnetic field to be constant and homogeneous.
The field splits the otherwise degenerate energy eigenstates
of the old Hamiltonian. The eigenstates of the new Hamiltonian (with magnetic field)
can be written down as:
\begin{align}
|\psi_{1}\rangle =&|11\rangle\nonumber\\
|\psi_{2}\rangle =&|1-1\rangle\nonumber\\
|\psi_{3}\rangle = &\frac{1}{\sqrt{2}}[|10\rangle+|00\rangle]\nonumber\\
|\psi_{4}\rangle =&\frac{1}{\sqrt{2}}[|10\rangle-|00\rangle]
\label{eq:eigenstates}.
\end{align}
which are linear combinations of the eigenstates of the old Hamiltonian(without magnetic field), viz. $|11\rangle$,$|1-1\rangle$, $|10\rangle$ and $|00\rangle$. \autoref{eq:eigenstates} clearly shows 
the mixing between one of the triplet $|10\rangle$ and singlet states $| 00\rangle$. 
The discussion up to this point is just a  recapitulation of  previous work \cite{michael2013,yang2011,filip2013}.

In the present treatment, we  take a leap forward to consider the actual scenario
which might be much more complicated. So, we  refrain  
from considering a homogeneous magnetic field. In the co-moving frame of heavy quarkonia,
moving in an inhomogeneous magnetic field,
the interaction Hamiltonian $H_{I}$ becomes solely time dependent as the magnetic field appears 
to be changing its magnitude and direction due to the inhomogeneity. Hence,
\begin{equation}
H_{I}(t)= -\omega(t)(S_{\hat{u}(t)Q}-S_{\hat{u}(t)\bar{Q}})~.
\label{eq:IntHamiltonian}
\end{equation}
Here $\omega=g\mu_QB$ is the Larmor frequency of the system and $S_{u}$ is the spin 
along the direction $\hat{u}$ making an angle $\theta$ with the z-axis and an azimuthal
angle $\phi$:
\begin{equation}
\hat{u}=\sin\theta \cos\phi~\hat{x}+\sin\theta \sin\phi~\hat{y}+\cos\theta~\hat{z} \label{direction}.
\end{equation}
Unlike the case with homogeneous magnetic field where the field is considered to be applied
in a fixed direction (say $z$), in the present circumstance (with inhomogeneous field) the 
field is changing its magnitude and direction. So, it is customary to introduce an arbitrary direction, $\hat{u}(t)$, along which the field would be aligned at any instant. It is worth noting that, in this
case, $H_I(t)$ and $S_{\hat{u}(t)}$ have common set of eigenvectors.

The interaction Hamiltonian $H_{I}(t)$ dictates the nature of the evolution of the 
energy eigenstates. Before we discuss the consequences of  time 
dependence of the interaction Hamiltonian, let us check whether the evolution of spin states
can really be adiabatic or not.

Instantaneous eigenstates of $H_I(t)$ are given by
\begin{align}
|\psi_{1}(t)\rangle &=|11\rangle^{\hat{u}(t)}~, 	\nonumber\\
|\psi_{2}(t)\rangle &=|1-1\rangle^{\hat{u}(t)}~,	\nonumber\\
|\psi_{3}(t)\rangle &=\frac{1}{\sqrt{2}}[|10\rangle^{\hat{u}(t)}+|00\rangle^{\hat{u}(t)}]~,	\nonumber\\
|\psi_{4}(t)\rangle &=\frac{1}{\sqrt{2}}[|10\rangle^{\hat{u}(t)}-|00\rangle^{\hat{u}(t)}]~.
\label{eq:instantaneous}
\end{align}
Here,
\begin{align}
|11\rangle^{\hat{u}(t)} &=|\uparrow\rangle^{\hat{u}(t)}_Q \otimes |\uparrow \rangle^{\hat{u}(t)}_{\bar{Q}}~,	\nonumber\\
|1-1\rangle^{\hat{u}(t)} &= |\downarrow\rangle^{\hat{u}(t)}_Q \otimes |\downarrow\rangle^{\hat{u}(t)}_{\bar{Q}}~,	\nonumber\\
|10\rangle^{\hat{u}(t)} &=  \frac{1}{\sqrt{2}}\left[|\uparrow\rangle^{\hat{u}(t)}_Q \otimes |\downarrow\rangle^{\hat{u}(t)}_{\bar{Q}}
+|\downarrow\rangle^{\hat{u}(t)}_Q\otimes |\uparrow\rangle^{\hat{u}(t)}_{\bar{Q}}\right]~,	\nonumber\\
|00\rangle^{\hat{u}(t)} &= \frac{1}{\sqrt{2}}\left[|\uparrow\rangle^{\hat{u}(t)}_Q\otimes |\downarrow\rangle^{\hat{u}(t)}_{\bar{Q}}
-|\downarrow\rangle^{\hat{u}(t)}_Q|\uparrow\rangle^{\hat{u}(t)}_{\bar{Q}}\right]~.
\label{eq:notation}
\end{align}
$|\uparrow\rangle^{\hat{u}(t)}_{Q/\bar{Q}}$ and $|\downarrow\rangle^{\hat{u}(t)}_{Q/\bar{Q}}$ are the up and down spins of the 
quark and antiquark along the direction $\hat{u}(t)$.

\begin{align}
|\uparrow\rangle^{\hat{u}(t)}_{Q/\bar{Q}}= & \left(\begin{array}{cc} e^{-i \phi(t)/2}\cos{\theta/2}
\\e^{i \phi(t)/2}\sin{\theta/2} \end{array}\right)\nonumber\\
|\downarrow\rangle^{\hat{u}(t)}_{Q/\bar{Q}} = &\left(\begin{array}{cc} e^{-i \phi(t)/2}\sin{\theta/2}
\\-e^{i \phi(t)/2}\cos{\theta/2} \end{array}\right).
\label{eq:updown}
\end{align}
The magnetic field here is considered to have inhomogeneity in the azimuthal plane only. 
For the evolution to be adiabatic, 
$\left|\frac{\langle\psi_m|\dot{\psi}_n\rangle}{E_m-E_n}\right|$ should be much less than $1$ \cite{amin2009,aharonov1987}
In the lab frame, the ratio comes out to be:
\begin{equation}
\left|\frac{\langle\psi_m|\dot{\psi}_n\rangle}{E_m-E_n}\right| = \left|\frac{\langle\psi_m|-
(\vec{v}.\vec{\nabla} )(\vec{\mu}.\vec{B})_{lab}|\psi_n\rangle}{E_m-E_n}\right|~.
\end{equation}
$E_m$ is the energy eigenvalue of the corresponding $m^{th}$ eigenstate $|\psi_m\rangle$.
As is reflected from the above expression, the ratio in this context depends 
on the velocity $\vec{v}$ of quarkonia as well as on the inhomogeneity of the magnetic field. 
Now, plugging in different energy eigenstates as given in \autoref{eq:instantaneous} we have 
$\left|\frac{\langle\psi_m|\dot{\psi}_n\rangle}{E_m-E_n}\right| = \frac{0}{0}$ for 
$m, n=1,2 ; m\neq n $ as $E_{1}$ and $E_{2}$ are the equal eigenvalues of the states $|\psi_1\rangle$
and $\\psi_2\rangle$ respectively. 
\begin{equation}
\left|\frac{\langle\psi_m|\dot{\psi}_n\rangle}{E_m-E_n}\right| = 0 \hspace{0.2cm} \text{for} m,n=3, 4; m\neq n~ .
\end{equation}
Other than these two cases, the value of the ratio is $\frac{\dot{\phi}\sin\theta}{2\omega} $. 
So, the ratio is determined through the degree of inhomogeneity $\dot{\phi}$, 
the angle $\theta$ and the Larmor frequency $\omega$. For a non-zero value of $\sin\theta$, 
the adiabatic evolution of the spin states is guaranteed if the Larmor frequency $\omega$ is high 
enough to cope with the changing direction of magnetic field experienced by the 
moving $q\bar{q}$ pair. This holds true for a very high value of magnetic field.
However, in heavy ion collisions, though a very 
strong magnetic field is supposed to be created in the beginning, it might persist for a very short 
duration. So, no matter how small the quantity $\dot{\phi}$ is, the adiabaticity is bound to be broken as 
$\omega \approx 0$ for vanishingly small magnetic field. 

\section*{Spin flipping transitions in weak field regime:}

As is already evident from the preceding discussion, the non adiabaticity
due to very small value of the magnetic field 
has an appreciable effect on the spin states of quarkonia. To quantify this effect, 
Schr\"odinger Equation for the spin states needs to be solved. 
Spin state of quarkonia at any instant is the linear combination of the 
instantaneous eigenstates, $|\psi_i(t)\rangle$, $i=1, 2, 3, 4$ given by \autoref{eq:instantaneous}:

\begin{equation}
|\Psi(t)\rangle = A_i | \psi_i(t)\rangle~.
\label{eq:lincom}
\end{equation}
 
\begin{figure*}
\centering
\begin{subfigure}[h!]{0.45\textwidth}
\includegraphics[width=\textwidth]{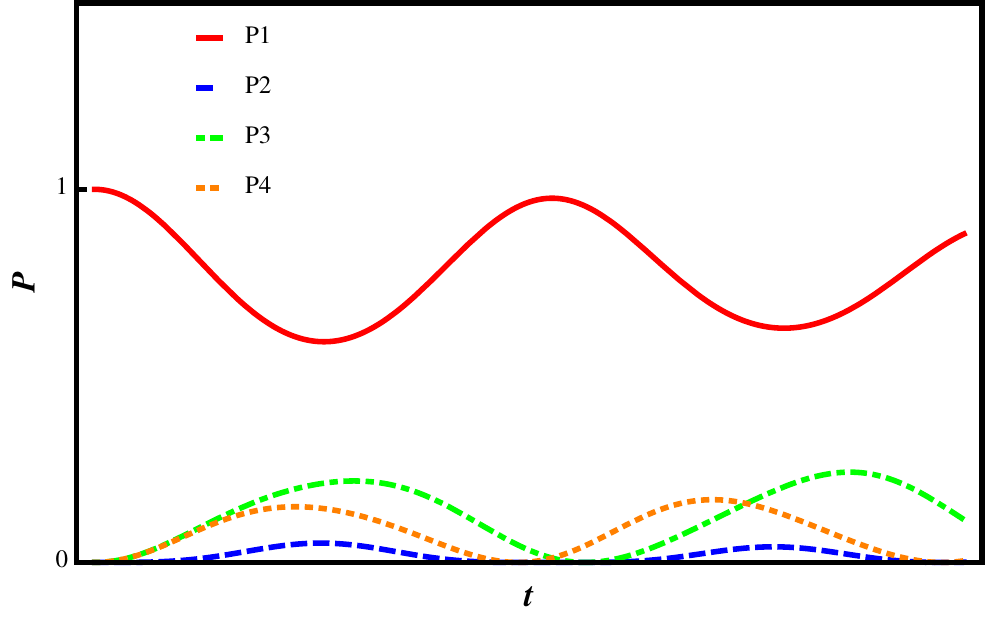}
\caption{}
\label{fig:theta04}
\end{subfigure}
\begin{subfigure}[h!]{0.45\textwidth}
\includegraphics[width=\textwidth]{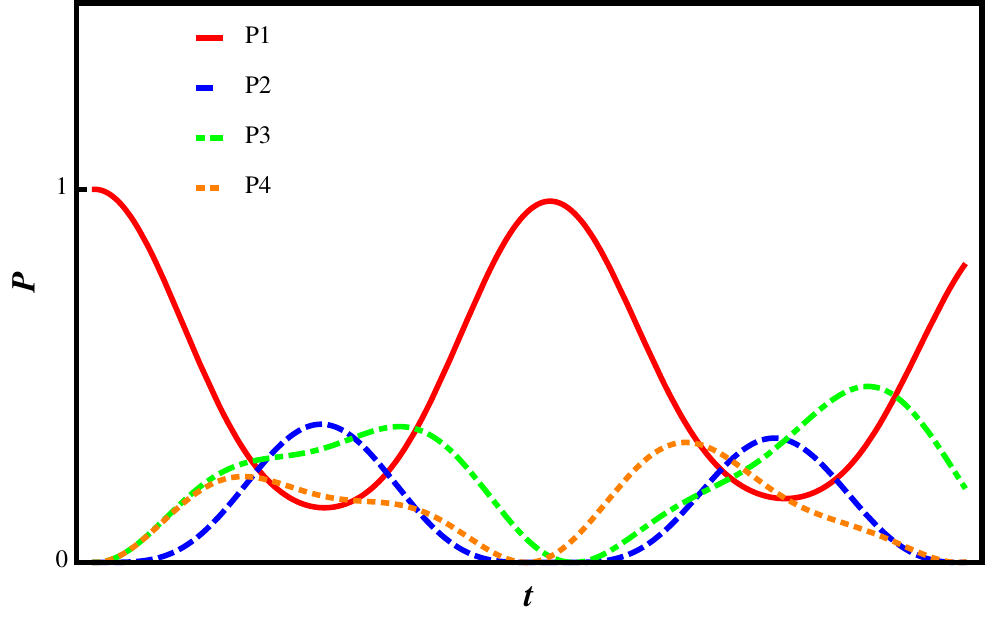}
\caption{}
\label{fig:theta09}
\end{subfigure}
\\
\vfill
\begin{subfigure}[h!]{0.45\textwidth}
\includegraphics[width=\textwidth]{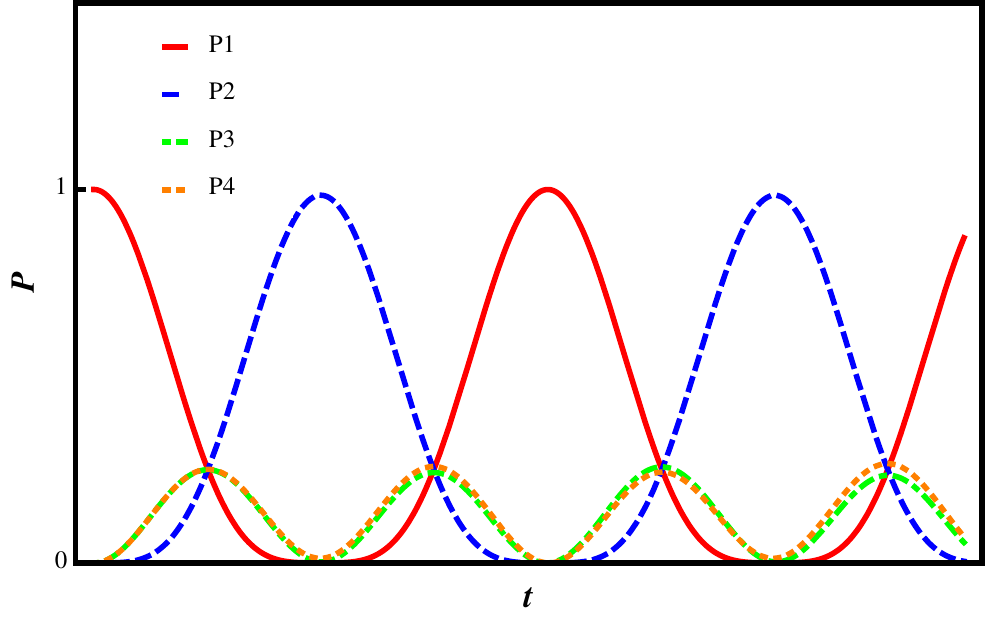}
\caption{}
\label{fig:theta16}
\end{subfigure}
\begin{subfigure}[h!]{0.45\textwidth}
\includegraphics[width=\textwidth]{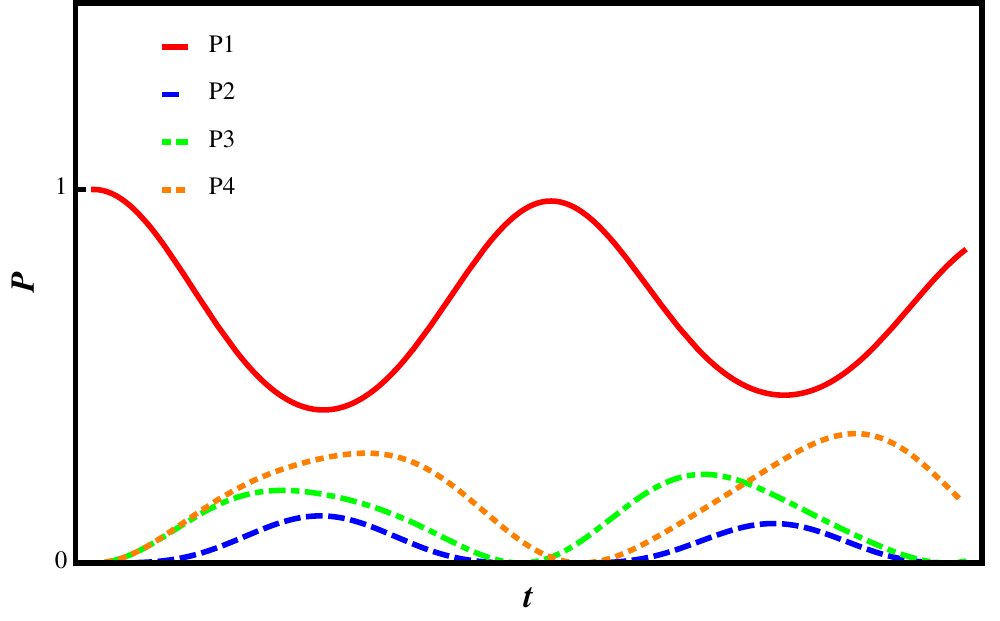}
\caption{}
\label{fig:theta25}
\end{subfigure}
\caption{Survival probability $P1$ and spin flipping transition probabilities $P2$, $P3$, $P4$ to other spin states $|\psi_2\rangle$, $|\psi_3\rangle$, $|\psi_4\rangle$ for different values of $\theta$ as in (a) $\theta = 0.5$, (b) $\theta = 0.9$, (c) $\theta = 1.6$, (d) $\theta = 2.5$ as a function of time.}
\label{fig:plots}
\end{figure*}

\autoref{eq:lincom} can be rewritten  
in terms of the basis $| 11 \rangle^{\hat{u}(t)}$, $|1-1\rangle^{\hat{u}(t)}$, $|10\rangle^{\hat{u}(t)}$ and $| 00\rangle^{\hat{u}(t)}$:
\begin{align}
|\Psi(t)\rangle
& \,
= C_1 
| 11\rangle^{\hat{u}(t)}
+ C_2
| 1-1\rangle^{\hat{u}(t)} \nonumber\\
& \,
+ C_3
| 10\rangle^{\hat{u}(t)}
+ C_4
| 00 \rangle^{\hat{u}(t)}
\label{eq:Psit}
\end{align}
The coefficients $C_i$'s and the states are time dependent. These states can be written as the direct 
product of individual up and down spin states of the two spin 1/2 particles (heavy quark and its corresponding
antiquark) bound to constitute a quarkonium. Time dependent Schr\"odinger Equation for $| \Psi(t)\rangle$ is,

\begin{equation}
i\frac{\partial}{\partial t}
| \Psi(t) \rangle
=
H_I(t) |\Psi(t)\rangle
\label{eq:SE}
\end{equation}
where $H_I(t)$ is the interaction Hamiltonian given by \autoref{eq:IntHamiltonian}.
Using \autoref{eq:Psit} and \autoref{eq:IntHamiltonian}, we arrive at:

\begin{align}
& i[\dot{C_1} |11\rangle^{\hat{u}(t)}
+ C_1|\dot{11}\rangle^{\hat{u}(t)} +\dot{C_2}|1-1\rangle^{\hat{u}(t)}\nonumber \\ 
&
+ C_2 |\dot{1-1}\rangle^{\hat{u}(t)}
+ \dot{C_3} |10\rangle^{\hat{u}(t)}
+ C_3| \dot{10}\rangle^{\hat{u}(t)} 	\nonumber \\
&
+ \dot{C_4}| 00\rangle^{\hat{u}(t)}
+ C_4 |\dot{00}\rangle^{\hat{u}(t)}] 	\nonumber\\
&
= C_1 H_I(t)|11\rangle^{\hat{u}(t)}
+ C_2 H_I(t)|1-1\rangle^{\hat{u}(t)} 	\nonumber \\
&
+ C_3 H_I(t)|10\rangle^{\hat{u}(t)}
+ C_4 H_I(t)|00\rangle^{\hat{u}(t)}
\label{eq:SEmodified}
\end{align}
Here, dots on top of the coefficients and the states signify their respective time derivatives. 
Taking inner product of \autoref{eq:SEmodified} with $\langle11|^{\hat{u}(t)}$, $\langle1-1|^{\hat{u}(t)}$, 
$\langle10|^{\hat{u}(t)}$ and $\langle00 |^{\hat{u}(t)}$ one at a time, we get four coupled first order differential equations for the 
coefficients $C_1$, $C_2$, $C_3$ and $C_4$ respectively. These equations take the following form when the magnetic field
intensity takes a negligibly small value and  azimuthal inhomogeneity red is considered upto the first order in the Taylor series of $\phi(t)$.

\begin{align}
\frac{dC_1}{dt} 
&=
iC_1\dot{\phi}\cos\theta+\frac{i}{\sqrt{2}}C_3\dot{\phi}\sin\theta ~,\\
\frac{dC_2}{dt}
&=
-iC_2\dot{\phi}\cos\theta+\frac{i}{\sqrt{2}}C_3\dot{\phi}\sin\theta~,\\
\frac{dC_3}{dt}
&=
\frac{i}{\sqrt{2}}C_1\dot{\phi}\sin\theta+
\frac{i}{\sqrt{2}}C_2\dot{\phi}\sin\theta+i\omega C_4~,\\
\frac{dC_4}{dt}
&=
i\omega C_3 ~.
\end{align}
One can plot the probability of getting a particular state given by its corresponding $\mid A_i(t)\mid^2$ 
as a function of time with various initial conditions and different values of Larmor 
frequency and  degree of inhomogeneity. 
Starting with the spin state $|\psi_1\rangle$, we  evaluate the survival probability $P_1$ and spin flipping transitions
$P_2$, $P_3$, $P_4$ to other three states $|\psi_2\rangle$, $|\psi_3\rangle$ and $|\psi_4\rangle$ respectively. In \autoref{fig:plots}, $P_1$, $P_2$,
$P_3$ and $P_4$ have been represented by red, blue(dashed), green(dot dashed) and orange(dotted) colours for different values of $\theta$. It is quite clear
from the plots that though we start with $|\psi_1\rangle$, all other spin states get mixed with a mixing probability which changes with time. This phenomenon also occurs even if we start with any other initial states($|\psi_2\rangle$, $|\psi_3\rangle$ and $|\psi_4\rangle$).

These plots confirm our previous assertion that if a rapidly decaying (with time) and
inhomogeneous magnetic field exists, the evolution of spin states of quarkonia becomes nonadiabatic in the regime 
where the strength of the field is very weak. We have shown that this nonadiabaticity results in a 
dynamical spin mixing among all possible spin states of quarkonia, no matter which initial state we
start with. However, the deconfined medium, once 
formed in HICs, might drag the magnetic field along with it and help it maintain an appreciably large value for 
quite some time \cite{ajit2017}. That being the scenario, the spin state evolution might have the possibility to be 
adiabatic and in turn, there will be no dynamical mixing occurring whatsoever. Hence,
the manifestation or absence of this dynamical mixing among all the spin sates of quarkonia, as presented 
in this paper, can be a good way to decide the actual nature, strength and  persistence of
the magnetic field before and after the formation of the QCD medium in HICs.

\bibliographystyle{apsrev4-1}
\bibliography{mfquarks}{}

\end{document}